\documentclass[twocolumn,prbrapid,superscriptaddress,showpacs]{revtex4-1}
\usepackage{amsmath,amssymb}
\usepackage{graphicx}
\usepackage{verbatim}
\usepackage{amsfonts}
\usepackage{dcolumn}
\usepackage{bm}
\usepackage{subfigure}
\usepackage{color}
\usepackage[colorlinks,citecolor=blue]{hyperref}

\begin{document}
\
\title{\large \bf Fermionic Symmetry Protected Topological Phase Induced by Interactions}

\author{Shang-Qiang Ning}
\affiliation{Institute for Advanced Study, Tsinghua University, Beijing, 100084, P. R. China}

\author{Hong-Chen Jiang}
\email{hongchen777@gmail.com}
\affiliation{Stanford Institute for Materials and Energy Sciences, SLAC National Accelerator Laboratory, 2575 Sand Hill Road, Menlo Park, CA 94025, USA}

\author{Zheng-Xin Liu}
\email{liuzxhk@gmail.com}
\affiliation{Institute for Advanced Study, Tsinghua University, Beijing, 100084, P. R. China}

\begin{abstract}
Strong interactions can give rise to new fermionic symmetry protected topological phases which have no analogs in free fermion systems. As an example, we have systematically studied a spinless fermion model with $U(1)$ charge conservation and time reversal symmetry on a three-leg ladder using density-matrix renormalization group. In the non-interacting limit, there are no topological phases. Turning on interactions, we found two gapped phases. One is trivial and is adiabatically connected to a band insulator, while another one is a nontrivial symmetry protected topological phase resulting from strong interactions.
\end{abstract}

\maketitle

\textbf{Introduction}. Gapped quantum states can be classified by their quantum entanglement\cite{ChenGuWen2010}. The long-range entangled states carry intrinsic topological orders\cite{Wen89, WenNiu90, Wen1990}, while the short-range entangled states are trivial and can be adiabatically connected to direct product states (or slater determinant states for fermionic systems). If the system has some symmetries, there will be more phases.

For instance, short-range entangled states without symmetry breaking can belong to different phases. Besides the trivial symmetric phases, there may exist symmetry protected topological (SPT) phases\cite{GuWen2009,Pollmann2010} which have nontrivial edge excitations. A typical example of bosonic interacting SPT phase is the $S=1$ Haldane phase\cite{HaldanePLA1983,HaldanePRL1983}, which is protected by spin rotational symmetry, or time reversal symmetry, or spatial inversion symmetry. Bosonic SPT phases with symmetry $G$ in $d$-dimension can be classified by the $(d+1)$th group cohomology $\mathcal H^{d+1}(G,U(1))$\cite{ChenGuLiuWen2011}.

SPT phases also exist in free fermion systems. Topological insulators\cite{KM0502, BZ0602, MB0706, FKM0703, QHZ0824} are well known SPT phases protected by $U(1)$ charge conservation and time reversal symmetry. Free fermion systems with different symmetry groups in different dimensions can be classified using homotopy theory, with 10 different classes of topological phases\cite{AltlandZirbauer1997, Schnyder-2008, Kitaev2009, Ryu-2010, Wen2012}. SPT phases may also exist in the presence of strong fermion-fermion interactions. However, the classification of the interacting fermionic SPT phases are usually different from those of free fermions. In 1D, interaction fermionic SPT phases can be classified by projective representations of the symmetry group\cite{ChenGuWen2011_1D, ChenGuWen2011_1Dfull}. In higher dimensions, the classification is more difficult and is partially described by the super-cohomology of the symmetry group\cite{GuWen2012}. Some examples of 2D interacting fermionic SPT phases are studied recently\cite{YaoRyu2012,Qi2012,RyuZhang2012}.  

An interesting question is what is the relation between the classification of SPT phases for the interacting and non-interaction systems. For bosonic systems (including spin systems), there are no nontrivial SPT phases without interaction. So all nontrivial Bosonic SPT phases are induced from interactions. In contrast, situations are quite different for fermionic systems. Naively speaking, strong interactions will reduce the number of SPT phases for fermions. For example, 1D free femion superconductors with time reversal symmetry have $\mathbb Z$ classes of topological phases, which reduces to $\mathbb Z_8$ under strong interactions\cite{FidkowskiKitaev2010, FidkowskiKitaev2011}. Another example is superconductors protected by $U(1)$ spin rotational symmetry and time reversal symmetry, where interactions reduce the classification of SPT phases from $\mathbb Z$ to $\mathbb Z_4$\cite{TangWen2012}. However, similar to bosonic systems, it is also possible that interactions can induce new SPT phases in fermionic systems. That is to say, some interacting SPT phases may have no analogs in free fermion systems. In this paper, we will illustrate this possibility through a concrete model in one dimension.  

Here we consider a spinless fermion model with $U(1)\rtimes Z_2^T$ symmetry, where $U(1)=\{e^{i\hat N\theta};\theta\in[0,2\pi)\}$ is the charge conservation symmetry and $Z_2^T=\{I,T\}$ is the time reversal symmetry with $Te^{i\hat N\theta}=e^{-i\hat N\theta}T$ and $T^2=1$. The spinless fermions can be interpreted as fully polarized electrons in a strong Zeeman field along $z$-direction, whereas the time reversal is defined as $\tilde T=e^{iS_x\pi}T$ with $\tilde T^2=1$. Without interaction, the classification of 1D SPT phases for $U(1)\rtimes Z_2^T$ symmetry is $\mathbb Z_1$\cite{Wen2012}, thus there is only one trivial band insulating phase. However, since $\mathcal H^2(U(1)\rtimes Z_2^T,U(1))=\mathbb Z_2$\cite{ChenGuLiuWen2011}, the symmetry group has two projective representations, indicating that there are two SPT phases under strong interactions. One is trivial and is adiabatically connected to the trivial band insulator. On the contrary, another one is nontrivial and has symmetry protected edge states, which cannot be connected to the trivial phase without closing the bulk gap if the symmetry is reserved. In the following we will explicitly construct the model and study the phase diagram using density matrix renormalization group. Generalization of our results to higher dimensions will also be discussed.

\begin{figure} 
\centering
\subfigure[The three-leg ladder]{
\includegraphics[width=3.3in]{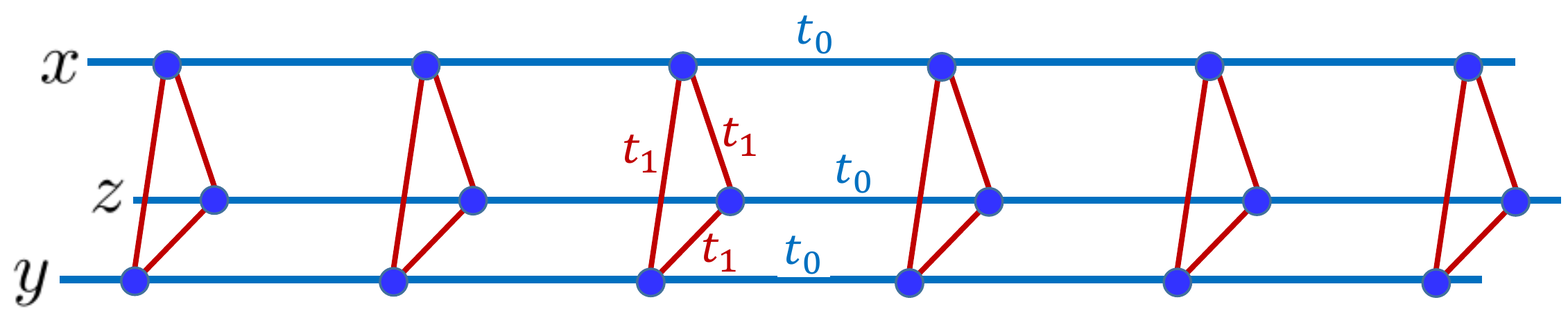}\label{fig:cartoon}}
\subfigure[Periodic boundary condition]{
\includegraphics[height=3cm ,width=4cm]{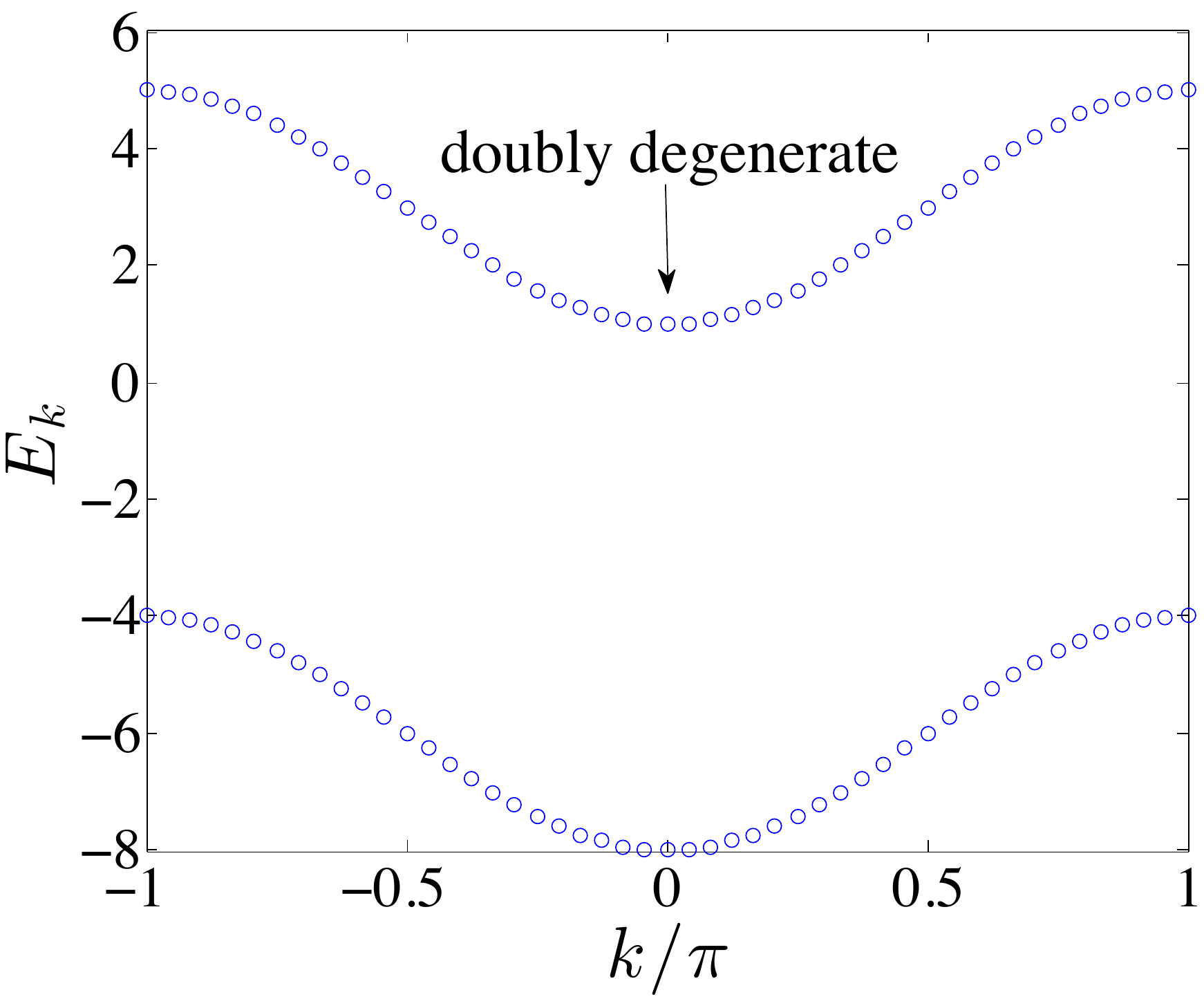}\label{freebulkenergy}}
\subfigure[Open boundary condition]{\includegraphics[height=3cm ,width=4cm]{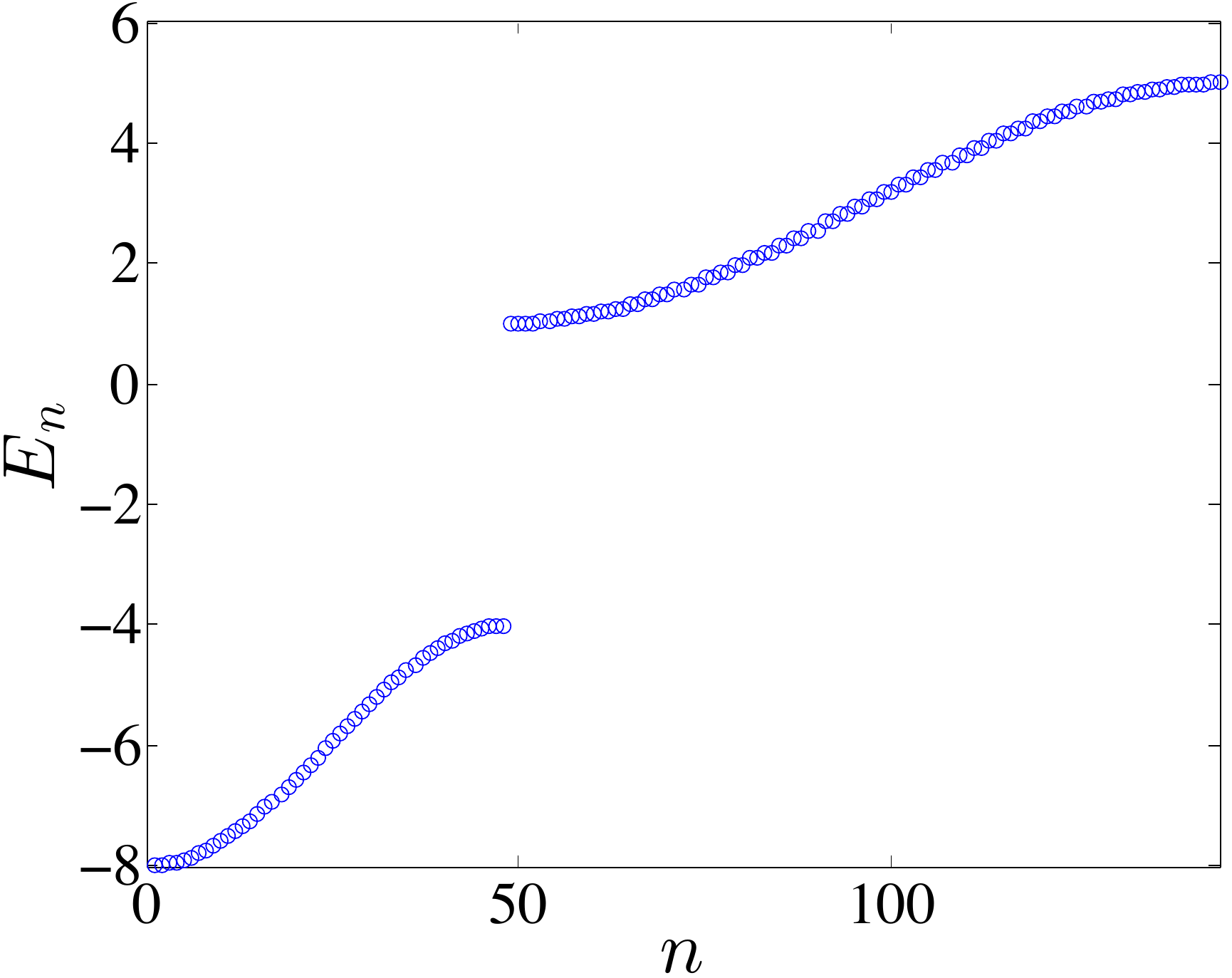}\label{freeopen}}
\caption{ (Color online) 
(a) Schematic picture of the spinless fermion model (\ref{Eq:ModelHam}) on a three-leg ladder; (b) Dispersion of the model with $t_1/t_0=3$ and $U=J=0$ under periodic boundary condition (here $N_r=48$). The upper band is doubly degenerate and the system is a band insulator at $1/3$-filling. If $t_1/t_0<{4\over3}$, the gap will close and the system becomes a metal; (c) Dispersion under open boundary condition. There are no zero modes. 
}\label{fig1}
\end{figure}

\textbf{The model}. We consider a spinless fermion model on a three-leg ladder with Hamiltonian
\begin{eqnarray}\label{eq:H}
H = H_{0}+H_{U}+H_{J}, \label{Eq:ModelHam}
\end{eqnarray}
where $H_0$ is the non-interacting Hamiltonian including the intra-chain hopping (with amplitude $t_0$) and inter-chain hopping (with amplitude $t_1$) terms (see Fig.~\ref{fig:cartoon} for detail)
\begin{eqnarray*}
H_0 &=& -\sum_{i}[t_0(c_{i,x}^\dagger c_{i+1,x}+c_{i,y}^\dagger c_{i+1,y}+c_{i,z}^\dagger c_{i+1,z}+{\rm h.c.} )  \\
&& + t_1(c_{i,x}^\dagger c_{i,y}+c_{i,y}^\dagger c_{i,z}+c_{i,z}^\dagger c_{i,x}+{\rm h.c.} ) +\lambda N_i],
\end{eqnarray*}
where $c_{i,x}$, $c_{i,y}$ and $c_{i,z}$ are anihilation operators of three species of spineless fermionons and $N_i=n_{i,x}+n_{i,y}+n_{i,z}$ is the total particle number at the $i$th rung. $H_U$ is the on-site Hubbard repulsive interaction
 \[
 H_U = U\sum_i(N_i-1)^2,
 \]  
and $H_J$ is a Heisenberg-like interaction
 \[
 H_J = J\sum_i \mathbf S_i\cdot \mathbf S_{i+1}\] with spin operators defined as $S^\alpha=\sum_{\beta,\gamma} i\varepsilon^{\alpha\beta\gamma}c_\beta^\dag c_\gamma$, where $\alpha,\beta,\gamma=x,y,z$. Under symmetry operations, the fermions vary in the following way
 \begin{eqnarray*}
&&U_\theta c_\alpha U_\theta^{-1}=c_\alpha e^{i\theta},\\
&&Tc_\alpha T^{-1} = - c_\alpha.
\end{eqnarray*}
It is obvious that the model (\ref{eq:H}) is invariant under the $U(1)\rtimes Z_2^T$ group.

\textit{Non-interacting limit.} We first study the free fermion model at ${1\over3}$-filling ({\it i.e.} there is one fermion at each rung in average) with $U=0, J=0$. According to the classification theory, there is only one gapped phase, {\it i.e.} the trivial band insulator. This can be easily seen from the bulk excitation spectrum. Under periodic boundary condition, the band structure is given as $E_{k}^{\rm v}=-2t_1-2t_0\cos(k),\ E_{k}^{\rm c}=t_1-2t_0\cos(k)$, where $E^{\rm v}$ is the valence band and $E^{\rm c}$ stands for the 2-fold-degenerate conducting bands. When $t_1 < {4\over3}t_0$, the system is a metal where both the valence band and the conducting bands are partially filled. When  $t_1 > {4\over3}t_0$, only the valence band is filled and the system becomes a band insulator (see Fig.(\ref{freebulkenergy})). The trivialness of the insulating phase can be reflected by the absence of zero edge modes under open boundary condition (see Fig.\ref{freeopen}). In the limit $t_1/t_0\to\infty$, the bands become flat, and the `charges' are localized at each rung in the ground state.

\textit{Strong interaction limit}. In the large $U$ limit, the system is deep in the Mott insulating phase and the `charges' are localized at each rung. In this case, the three species of fermions on each rung effectively act as the three components of a $S=1$ spin. If $t_1>0$, the spin degrees of freedom will be fixed and the ground state is unique. 

\begin{figure}
\centering
\includegraphics[width=3.3in]{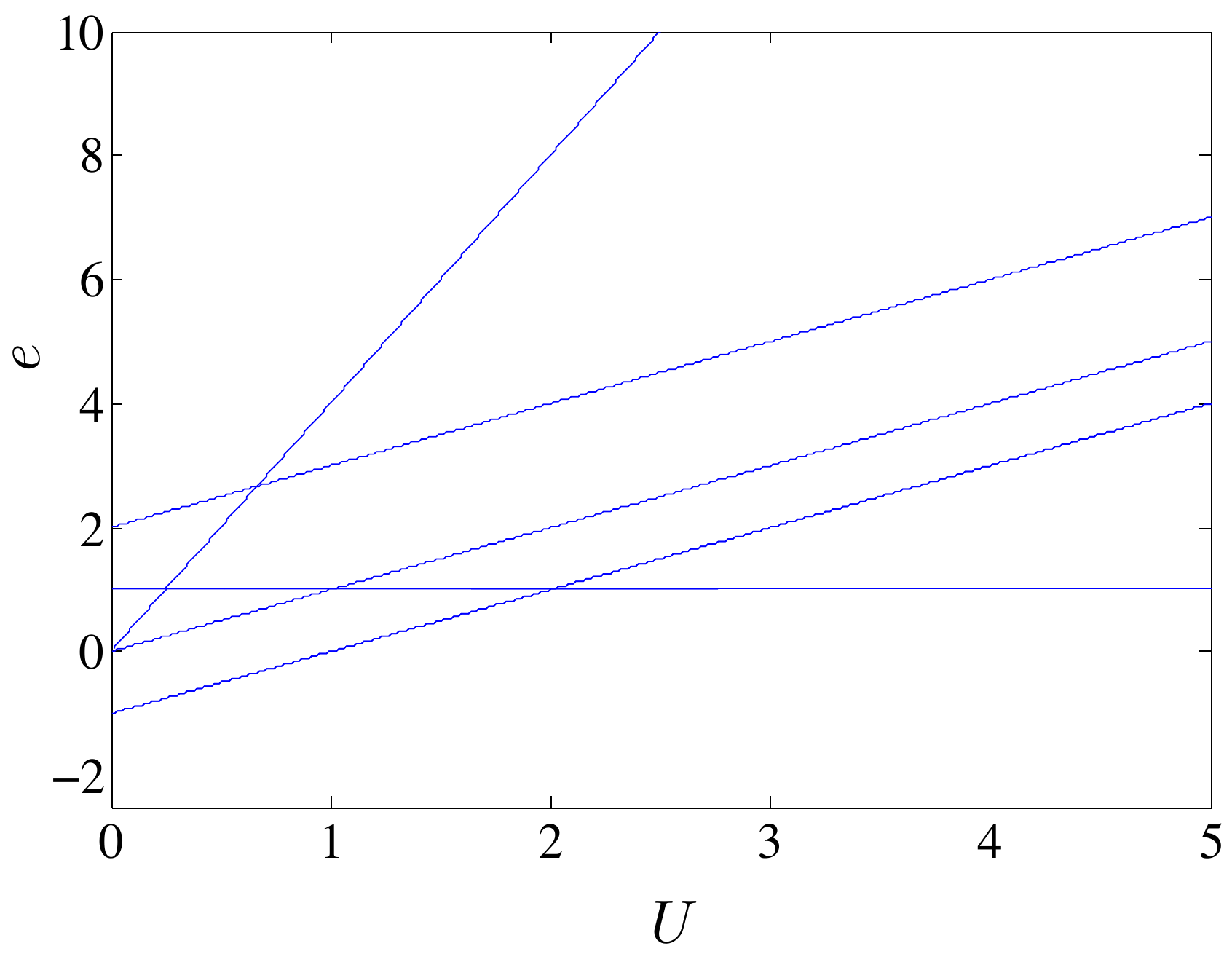}
\caption{\footnotesize  (Color online) Energy spectrum of a single rung (with $t_0=0$ and $t_1=1$) as a function of Hubbard repulsive interaction $U$. The red line denotes ground state energy, and the blue lines denote excited states.  The gap remains finite when $U$ varies from $0$ to $+\infty$.}
\label{Ufromzerotoinfinty}
\end{figure}

Suppose we increase the on-site repulsive interaction $U$ from 0 to infinity in the limit $t_1/t_0=\infty$ ($i.e.\ t_1=1,\ t_0=0$), where the system becomes decoupled rungs. The `charges' are always localized at each rung and there is a finite gap above the ground state for all values of $U$ (see Fig.~\ref{Ufromzerotoinfinty}). That is to say, the band insulator at $U=0$ and the Mott insulator at large $U$ are adiabatically connected and belong to the same trivial phase. {
This result also holds for nonzero $t_0$ with $t_0 < {3\over4}t_1$.}

In the Mott region (when $U$ is large enough), the $S=1$ spin degrees of freedom dominate the low energy physics. If we couple the rungs via strong Heisenberg-like interaction $H_J$, the system will be in another gapped phase --- the fermionic Haldane insulating phase. The difference between the fermionic Haldane insulator and the bosonic Haldane phase in spin-1 chains\cite{HaldanePLA1983,HaldanePRL1983} or extended Bose-Hubbard model\cite{Torre06,Berg08} is that the former has finite fermionic charge fluctuations. The existence of the fermionic Haldane insulator is guaranteed by the fact that the symmetry group $U(1)\rtimes Z_2^T$ has a nontrivial projective representation(see the supplementary material) .

\textbf{Numerical studies.} Above we have analyzed some limits of the spinless fermion model (\ref{eq:H}) and illustrated that there are three phases, one of them is induced by strong interactions. Now we will determine the ground state phase diagram and properties of the system by extensive and accurate density matrix renormalization group\cite{White1992DMRG} (DMRG) simulations. In the calculation, we consider a system with total number of sites $N=3N_r$, where $N_r$ is the number of rungs (up to 96) and $3$ is the number of legs. For simplicity, we will set $t_0=1$ as the energy unit and use open boundary condition. We keep up to $m=1536$ states in the DMRG block with around $10$ sweeps to get converged results. The truncation error is of the oder $10^{-6}$ or smaller.

\begin{figure}[t]
\centerline{
    \includegraphics[width=3.5in] {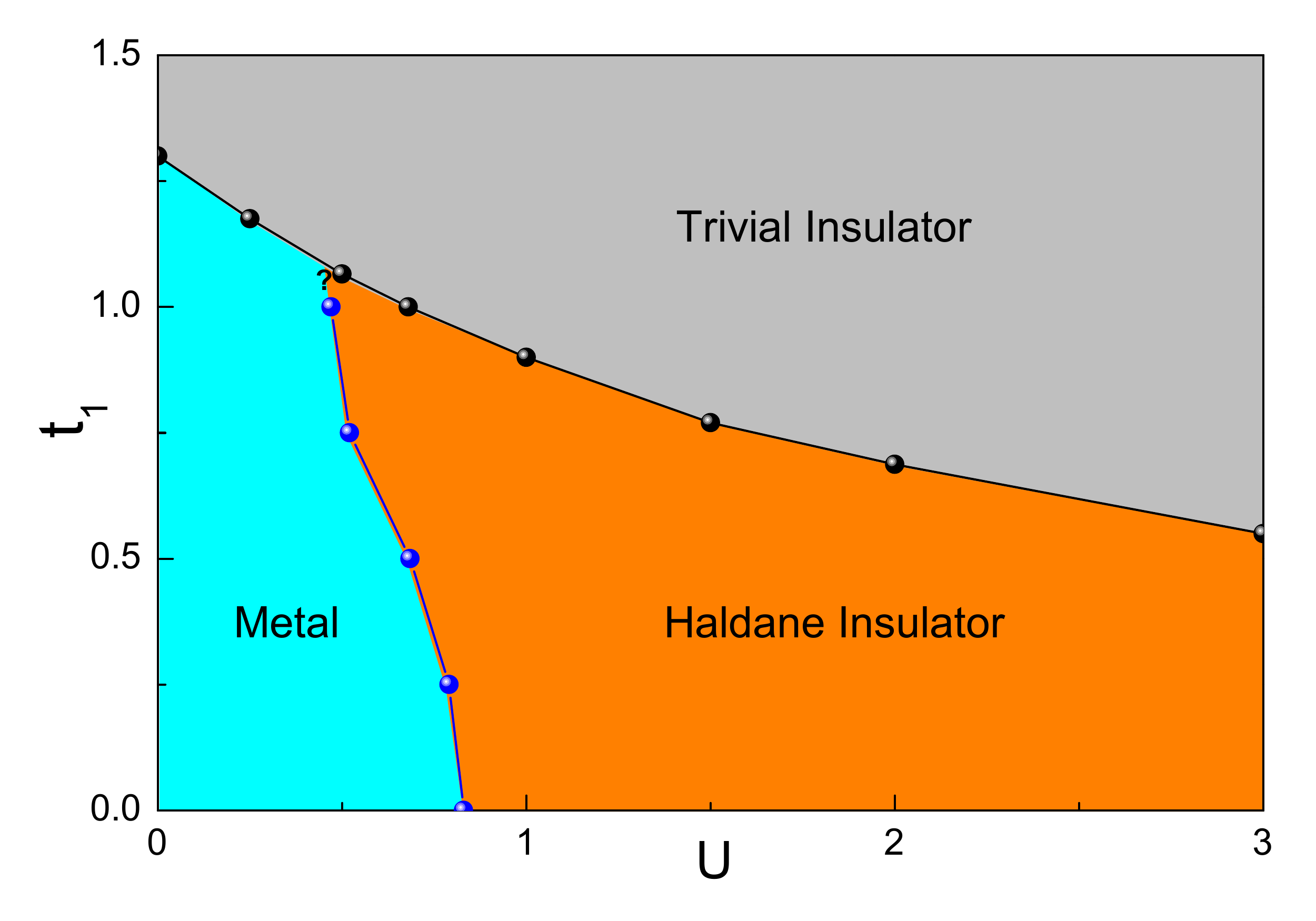}
    }
\caption{(Color online) Phase diagram for the spinless fermion model [Eq.(\ref{Eq:ModelHam})] at $1/3$-filling with $J=0.5$ and $N_r=48$, as determined by DMRG calculations. Varying parameters $t_1$ and $U$, three different phases are found: metal, trivial insulator and Haldane insulator. {
The question mark indicates a possible tricritical point whose precise location is difficult to determine numerically.}}
\label{Fig:PhaseDiagram}
\end{figure}

\begin{figure}[t]
\centerline{
    \includegraphics[width=3.2in] {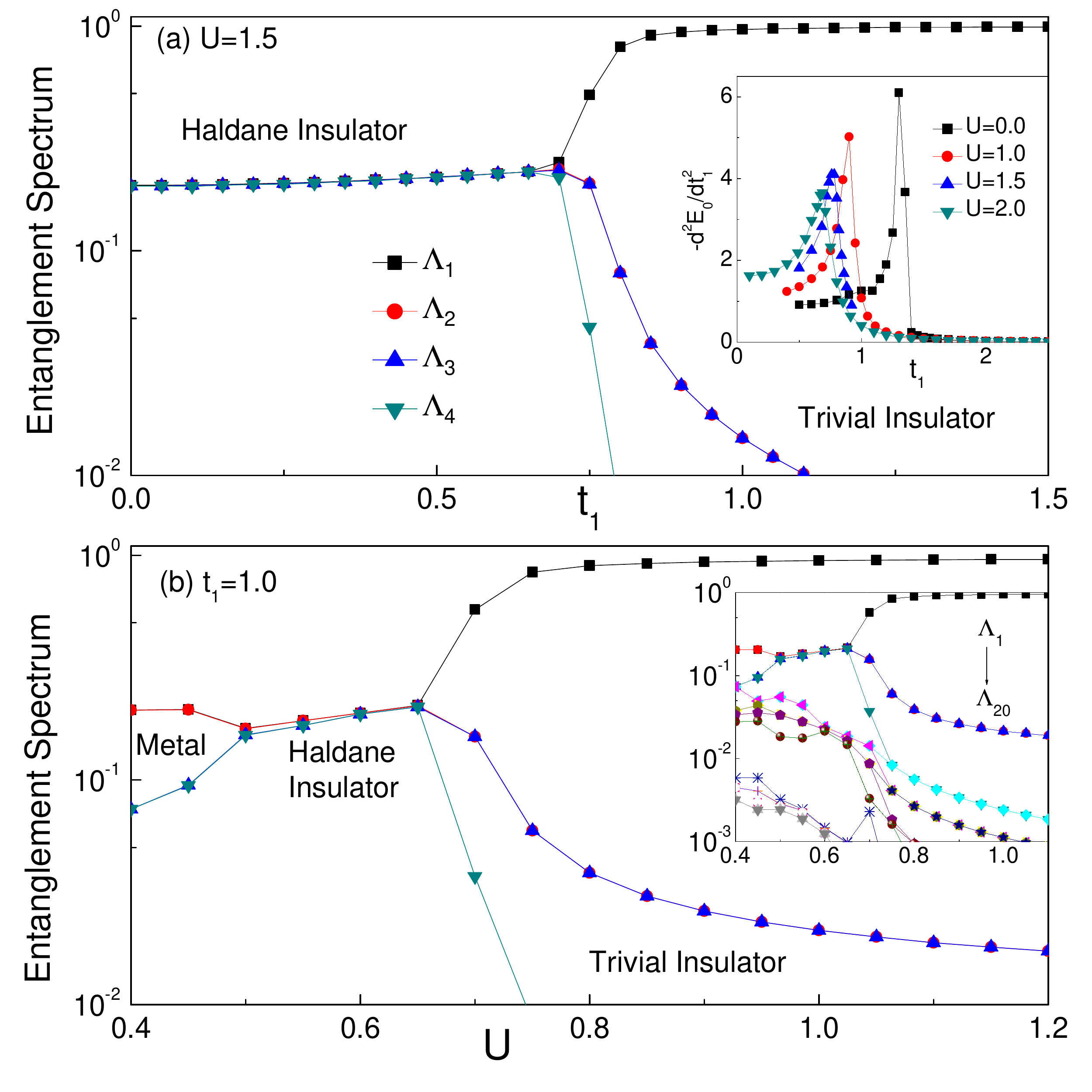}
    }
\caption{(Color online) Evolution of entanglement spectrum (with the largest 4 weights $\Lambda_1\sim \Lambda_{4}$) for the spinless fermion model in Eq.(\ref{Eq:ModelHam}), at (a) $U=1.5$ as a function of $t_1$, and (b) $t_1=1.0$ as a function of $U$. Inset in (a): Second derivative of the ground state energy density $-\frac{\partial ^2E_0}{\partial t_1^2}$ with different $U$. {
Inset in (b): Entanglement spectrum $\Lambda_1\sim \Lambda_{20}$ for $U=1.5$ as a function of $t_1=1.0$. Here $N_r=48$}.} 
\label{Fig:ESpectrum}
\end{figure}

 The main result is illustrated in the phase diagram shown in Fig.\ref{Fig:PhaseDiagram} at filling $\rho=\frac{1}{3}$ and $J=0.5$. Changing the coupling parameters $t_1$ and $U$, three different phases are found, including a gapless metal phase and two gapped phases --- the trivial insulator and Haldane insulator.  The Haldane insulator is a nontrivial SPT phase, which has degenerate edge states under open boundary condition. Numerically, the existence of the edge modes can be identified by the real-space entanglement spectrum (ES)\cite{LiHaldane2008, Pollmann2010}, {
 since the edge modes respect projective representation of the symmetry group and the degeneracy of the entanglement spectrum equals to the dimension of the irreducible projective representations}. The ES is defined as the set of eigenvalues of the reduced density matrix $\rho_A=\rm{Tr}_B|\Psi\rangle\langle\Psi|$, with $A$ being a subsystem({
 {\it e.g.} the left half part of the ladder}) and $B$ the remainder of the system, and $|\Psi\rangle$ is  the ground state wavefunction of whole system. As shown in Fig.\ref{Fig:ESpectrum}, the largest weight of ES in the Haldane phase is 4-fold degenerate, and this degeneracy is associated with the 2-fold degenerate edge states\footnote{For an infinite-long-ladder system, the degeneracy of ES should be 2-fold in the Haldane phase. However, for a finite ladder, the low energy edge states have weak coupling and results in entanglement between the edge states. Therefore, the ES have an extra 2-fold degeneracy. The same thing happens in the metal phase, where the low energy degrees of freedom on the edge may give rise to accidental 2-fold degeneracy of the ES.}. On the contrary, there is no such degeneracy in the ES of the trivial insulating phase and the metal phase.

\begin{figure}[t]
\centerline{
    \includegraphics[width=3.5in] {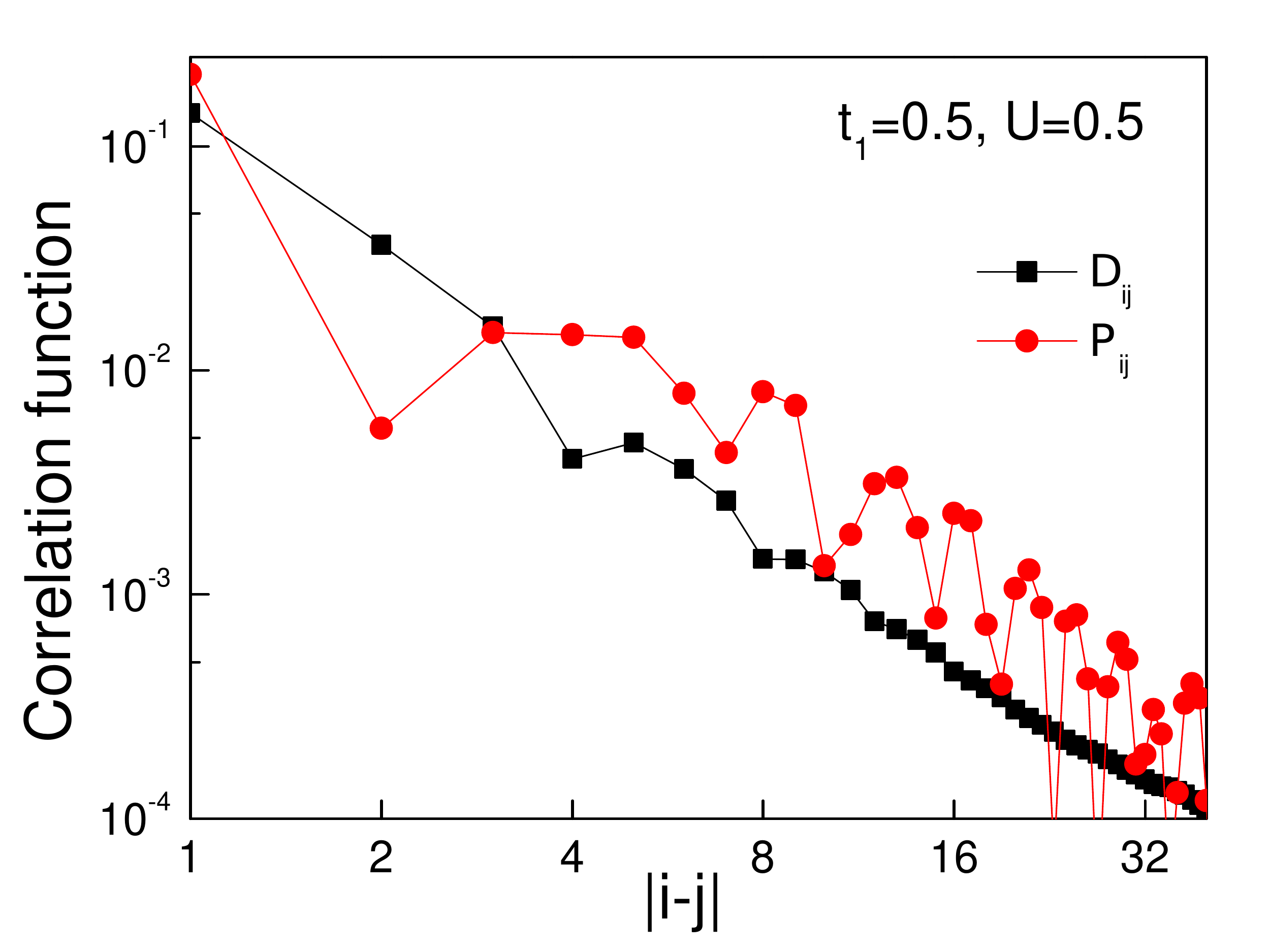}
    }
\caption{(Color online) Log-log plot of the density-density $D_{ij}$ and spin-spin $P_{ij}$ correlation functions for $t_1=0.5$, $U=0.5$. The $D_{ij}$ is power law decaying. The $P_{ij}$ is strongly fluctuating, and decays in power law on average. Here $N_r=96$.}
\label{Fig:MetalCorr}
\end{figure}

For small inter-chain hopping $t_1$ and weak interaction $U$, the metal phase is stable. In this phase, both the valence band and conduction bands are partially filled, and the system remains gapless. Both the fermion density-density correlation function $D_{ij}=\langle N_i N_j\rangle-\langle N_i\rangle\langle N_j\rangle$ (here $i$ and $j$ are rung indices) and spin-spin correlation function $P_{ij}=\langle S_i^z S_j^z\rangle - \langle S_i^z\rangle \langle S_j^z\rangle$ decay as power-law (see Fig.~\ref{Fig:MetalCorr}). 

\begin{figure}[t]
\centerline{
    \includegraphics[width=3.2in] {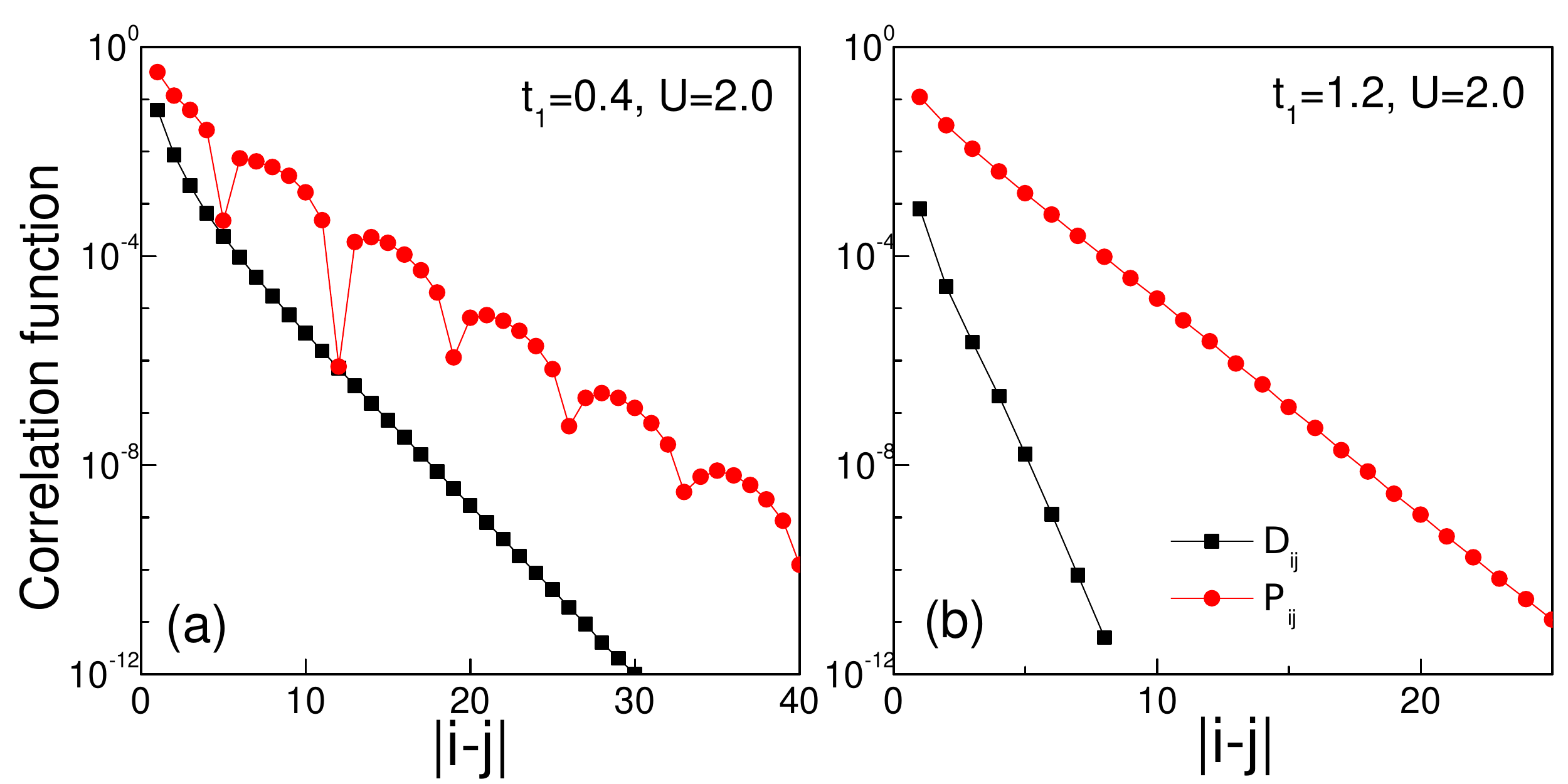}
    }
\caption{(Color online) Log-linear plot of density-density $D_{ij}$ and spin-spin $P_{ij}$
correlation functions for (a) $t_1=0.4$, $U=2.0$, and (b) $t_1=1.2$, $U=2.0$, respectively. Here $N_r=96$. }\label{Fig:Correlation}
\end{figure}

With the increase of inter-chain hopping $t_1$, the metal phase becomes shrinked and eventually gives way to trivial insulating phase. For weak interaction $U$, this trivial insulating phase is nothing but a band insulator, where the lower valence band is fully filled and the upper conduction bands are empty. For strong interaction $U$, the  system
becomes a Mott insulator which is adiabatically connected to the band insulator. In this trivial phase, all the excitations are gapped in the bulk. Therefore,  both the density-density $D_{ij}$ and spin-spin $P_{ij}$ correlation functions  decay exponentially, as seen in Fig.\ref{Fig:Correlation}(b). Although symmetry is unbroken in this phase, there are no protected edge modes.

For moderate inter-chain hopping $t_1$, the ground state of the system becomes the nontrivial Haldane insulator, when the interaction $U$ is strong enough. Similar with the trivial insulator, all the excitations in the bulk are gapped, with exponentially decaying density-density $D_{ij}$ and spin-spin $P_{ij}$ correlation functions [see Fig.\ref{Fig:Correlation}(a)]. However, as mentioned, the Haldane phase has nontrivial gapless edge modes protected by the $U(1)\rtimes Z_2^T$ symmetry. The Haldane insulator has no analog in free fermion models and is purely a consequence of strong interactions, especially the $J$-term interaction. If we set $J=0$, then the Haldane insulating phase will disappear. {
On the other hand, if we fix $U$ at a very large number, then if $t_1=0$, the system is gapless  and it is expected (but very difficult to verify numerically) that arbitrarily small $J>0$ will drive the system to the Haldane phase; if $t_1\neq0$, then the critical $J_c$ will be larger than 0.}

 The phase transition between the metal and trivial insulator, as well as the transition between the Haldane insulator and trivial insulator are second order. Therefore, the accurate phase boundaries can be directly determined by the second order derivative of the ground state energy density $-\frac{\partial^2E_0}{\partial t_1^2}$, as seen in the inset of Fig.\ref{Fig:ESpectrum}. The transition points obtained from the energy derivative are consistent with that obtained from the change of degeneracy of the ES [see Fig.~\ref{Fig:ESpectrum}(a) for the data at $U=1.5$]. By contrast, the phase transition between the metal and Haldane insulator seems Kosterlitz-Thouless like since the first and second order energy derivatives are smooth at the vicinity of the phase boundary. In this case the phase boundary is solely determined by the change of the degeneracy of ES [see Fig.\ref{Fig:ESpectrum}(b)].

\textbf{Conclusion and discussion.} In summary, we have systematically studied the one-dimensional spinless fermion ladder model (\ref{eq:H}) and shown that strong interactions can give rise to a new SPT phase. Without interaction, the model has only one band insulating phase. Turning on strong interactions, there will be two Mott insulating phases, one is adiabatically connected to the band insulator, while another one is nontrivial and has symmetry protected gapless edge modes. 

In the Mott limit $U=+\infty$, the charge degrees of freedom are completely frozen and the system reduces to a spin model with time reversal symmetry $Z_2^T$.  In this limit, the nontrivial SPT phase becomes the spin-1 Haldane phase. At finite $U$, the nontrivial SPT phase has similar spin dynamics but with nonzero charge fluctuations. In other words, the system contains fermionic charge excitations, although they exist at a relatively high energy. In this sense, the fermionic Haldane insulator at finite-$U$ is different from the pure bosonic Haldane phase. The stability of the fermionic Haldane insulator against charge fluctuations is protected by the nontrivial projective representation of the symmetry group $U(1)\rtimes Z_2^T$.  Notice that the Haldane insulating phase has also been discussed in two-legged  spin-1/2 fermion ladder models with Hubbard interactions\cite{Lecheminant10,Kawakami14,Rosch07,Pollmann14}. However, comparing with our spineless fermion model there is a subtle difference between the symmetry groups. For spin-1/2 fermions $T^2=P_f$ where $P_f$ is the fermion parity. As a result, the symmetry group is $G_-^-(U,T)$\cite{Wen2012} instead of $G_+^-(U,T)=U(1)\rtimes Z_2^T$. Since the group $G_-^-(U,T)$ has NO nontrivial projective representations, the Haldane insulating `phase' in the spin-1/2 fermionic ladder model is not stable against charge fluctuations and can be adiabatically connected to the band insulatior unless the model has extra inter-chain reflection symmetry\cite{Rosch07,Pollmann14}.

 Since $\mathcal H^2(U(1)\rtimes Z_2^T, U(1))=\mathcal H^2(Z_2^T,U(1))=\mathbb Z_2$, the interacting fermionic SPT phase corresponds to the bosonic SPT phase in the extreme Mott limit. This correspondence provides a method for constructing fermionic SPT phases in higher dimensions. For example, in three spatial dimensions, there are no nontrivial free fermion SPT phases with the same symmetry group $U(1)\rtimes Z_2^T$\cite{Wen2012},  but in the extreme Mott limit ($U\to\infty$) the $Z_2^T$ symmetry alone can protect three nontrivial bosonic SPT phases\cite{SenthilPRX13}. If these phases can survive under the charge fluctuations at finite $U$, then we will be able to obtain three interacting fermionic SPT phase which can not be realized in free fermion systems  (a similar symmetry,  $i.e.$, the combination of $U(1)$ and $Z_2^T$ with $T^2=P_f$ has been discussed in 3D in Ref.~\onlinecite{WangScience14}).  
In two dimensions, $U(1)\rtimes Z_2^T$ can also protect one nontrivial SPT phase \cite{LuVishwanath2012, LuVishwanath2014} which can not be realized without interactions.\cite{Wen2012} However, the realization is quite different from that in 1D, since this nontrivial phase can not be protected by the $Z_2^T$ symmetry alone (in the extreme Mott limit). Instead, the nontrivial SPT phase can be understood as a bosonic topological Mott insulator\cite{ LiuGuWen14} of a molecule system where each molecule is a bound state of even number of fermions.  

We thank Xiao-Gang Wen for initiating discussions and very helpful comments. We also thank Zheng-Cheng Gu and Xiong-Jun Liu for helpful discussions.  Work by ZXL and SQN were supported by NSFC 11204149 and Tsinghua University Initiative Scientific Research Program. Work by HCJ was supported by the Department of Energy,  Office of Science, Basic Energy Sciences, Materials Sciences and Engineering Division, under Contract DE-AC02-76SF00515.  We also acknowledge computing support from the Center for Scientific Computing at the CNSI and MRL: an NSF MRSEC (DMR-1121053) and NSF CNS-0960316.

\appendix
\section{Projective Representation of $U(1)\rtimes Z_2^T$ group}\label{sec:App}

One-dimensional SPT phases are classified by the second group cohomology $\mathcal H^2(G,U(1))$ of the symmetry group $G$. Mathematically, each element in the second group cohomology corresponds to a projective representation of the group $G$. That is to say, 1D SPT phases (either bosonic or fermionic) have one-to-one correspondence to the projective representations of the symmetry group. 

A representation of group $G$ is projective if the representation matrices satisfy the following relation:
\begin{equation} \label{Projective}
M(g_1)M(g_2)=e^{i\theta(g_1,g_2)}M(g_1g_2), 
\end{equation}
where the phase factor $\theta(g_1,g_2)$ is a function of two group elements $g_1, g_2$. Since (\ref{Projective}) holds for all group elements, it can be shown that the phase factor satisfy the following condition according to associativity of the multiplication:
\begin{equation} 
e^{i\theta(g_1,g_2g_3)}e^{i\theta(g_2,g_3)}=e^{i\theta(g_1,g_2)}e^{i\theta(g_1g_2,g_3)}.
\end{equation} 
On the other hand, two projective representations $\tilde M(g)$  and $M(g)$ are equivalent if they are related by a gauge transformation $\tilde M(g)=M(g)e^{i\varphi(g)}$ such that $e^{i\theta'(g_1,g_2)}=e^{i\theta(g_1,g_2)} {e^{i\varphi(g_1)}e^{i\varphi(g_2)}\over e^{i\varphi(g_1g_2)}}$. If the phase factor can be transformed into $e^{i\theta(g_1, g_2)}=1$ for all group elements under a gauge transformation, then the projective representation is the usual linear representation, $i.e.$, the trivial projective representation. 

Now we consider the symmetry group $U(1)\rtimes Z_2^T$. Since $\mathcal H^2(U(1)\rtimes Z_2^T, U(1))=\mathbb Z_2$ \cite{ChenGuLiuWen2011}, there are two classes of projective representations of the symmetry group $U(1)\rtimes Z_2^T$. The first one is a trivial one-dimensional linear representation, 
\begin{equation}
M(U_\theta)=e^{in\theta}, M(T)K=e^{i\phi}K, \ \ n\in\mathbb Z,
\end{equation} 
where $K$ is the anti-linear operator.
The second one is a two-dimensional nontrivial projective representation and can be chosen as
\begin{eqnarray} \label{2dProj}
M(U_{\theta})=\left(\begin{array}{cc}e^{in\theta}&0 \\ 0&e^{in\theta}\end{array}\right), M(T)K=\left(\begin{array}{cc} 0&-1\\1&0\end{array}\right)K, \ \ n\in\mathbb Z.\nonumber\\
\end{eqnarray} 
One can check that $M(U_{-\theta})M(T)K=M(T)KM(U_\theta)$ and $[M(T)K]^2=-1$.  Comparing to the multiplication of group elements $T^2=1$, the minus sign in the multiplication of representation $[M(T)K]^2=-1$ can not be gauged away. This nontrivial projective representation protects the 2-fold degenerate edge states in the Haldane insulating phase of the model (1) in the main text.

\bibliography{Liuzx}

\end{document}